%
%

\ifx\mnmacrosloaded\undefined \input mn \fi



\def\Mpc{\,\hbox{Mpc}}
\def\ergs{\,\hbox{erg}\,\hbox{s}^{-1}}

\def\cm{\,\hbox{cm}}

\def\kms{\,\hbox{km}\,\hbox{s}^{-1}}


\def\cv{{\rm c_v}}              
\def\smin{s_{\rm min}}          
\def\zmax{z_{\rm max}}          


\overfullrule=0pt

\input psfig


\begintopmatter

  \title{The Entropy-Driven X-ray Evolution of Galaxy Clusters}
  \author{Richard G. Bower,$^1$}
  \affiliation{$^1$ Dept.\ of Physics, University of Durham, South Road, Durham, 
	DH1 3LE, U.~K. (e-mail: R.G.Bower@durham.ac.uk)}
  \shortauthor{Richard~G.~Bower}
  \shorttitle{The Entropy-Driven Evolution of Clusters}

  \abstract{Observations of the evolution of the galaxy cluster X-ray 
	luminosity function suggest that the entropy of the intra-cluster 
	medium plays a significant role in determining the development of 
	cluster X-ray properties. I present a theoretical framework 
	in which the evolution of the entropy of the central intra-cluster 
	gas is explicitly taken into account. 
	The aim of this work is to develop a theoretical context 
	within which steadily improving measurements of the X-ray
	luminosities and temperatures of distant galaxy clusters can
	be interpreted. I discuss the possible range of entropy evolution
	parameters and relate these to the physical processes heating and 
	cooling the intra-cluster medium. The practical application
	of this work is demonstrated by combining currently available 
	evolutionary constraints on the X-ray luminosity function and the 
	luminosity--temperature correlation to determine the best-fitting
	model parameters.}

\maketitle

\section{Introduction}

A precise determination of the evolution of X-ray properties of clusters of
galaxies will shortly be available. In particular, the X-ray luminosity
function (XLF) and the luminosity--temperature (L-T) correlation are 
open to accurate measurement through, in the first case, serendipitous 
imaging surveys and, in the second, targeted spectral observations of 
known X-ray clusters. The aim of this paper
is to develop a theoretical framework within which these data may be placed. 

The first predictions for the evolution of cluster X-ray properties 
(eg., Kaiser, 1986) were based on the assumption of self-similar evolution
of both the cluster gravitational potential and the intra-cluster 
medium (ICM). Relatively few conditions need to be met in order
for the evolution of the dark-matter component to satisfy this
type of evolution. As such, these models are more generally applicable
than those that attempt to determine the distribution
of cluster masses explicitly (eg., Press \& Schechter, 1978). However,
whereas the Press-Schechter scheme can be used to predict the distribution 
--- and even conditional distributions --- of cluster properties,
the self-similar model provides only a scaling of cluster properties
between epochs. 

For realistic primordial density fluctuation spectra, application of 
the self-similar assumption to both cluster components predicts  
that the XLF evolution is strongly positive (ie., an 
increasing number of X-ray luminous clusters with increasing 
redshift or look-back time). This has, however, proved contradictory to the 
observational evidence: the present debate centres on whether
the X-ray luminosity function is non-evolving or whether it falls with
redshift (eg., Edge et al., 1990, Henry et al., 1992, Rosati et al., 1995,
Castander et al., 1995, Nichol et al., 1996). 

In order to solve this problem, Kaiser (1991) and Evrard \& Henry 
(1991, EH) proposed that the ICM was initially hot even before 
falling into the cluster. The early heating of the gas gives it an
initial entropy. Shock heating during cluster collapse may increase
the present-epoch entropy above this value, but cannot decrease
it. A fall in the entropy can only be achieved by radiative cooling
of the gas --- this occurs by recombination or Bremsstrahlung 
radiation, or through inverse Compton scattering of microwave 
background photons (cf., Padmanabhan, 1995). If the initial heat input
occurs at low enough redshift, these processes are inefficient
outside virialised regions, and the gas retains a memory of its
initial entropy as it clumps together to form larger and larger mass
units. Thus the ICM is imprinted with a minimum entropy and hence
a maximum density to which the core gas can be compressed without
greatly increasing the central temperature and pressure. 
The papers referenced above showed that the evolution of the XLF
was modified in the desired sense. However, this particular physical 
scenario creates only a single model. With steadily improving observations
of distant clusters, it is becoming possible to test between
a whole family of models that span these extremes, and even go beyond.
EH suggest the general form $L_x \propto (1+z)^s M^p$,
but only briefly explore the physical significance of their parameters
$s$ and $p$. 

In this paper, I present a more flexible scheme for including the effects of 
entropy. The new approach has the advantage that the evolutionary parameters 
may be directly interpreted in terms of the physical processes heating, 
pre-heating and cooling the ICM. Focusing the discussion on the evolution of
entropy automatically separates the contributions of the gravitational
evolution of the cluster (which results in adiabatic compression
of the ICM) and the specifically gas-phase phenomena, such as shock
heating and radiative cooling, which alter the adiabat on which the 
gas lies. The model thus parameterises the evolution of clusters in such 
a way as to provide a continuum spanning between the self-similar model 
and the constant entropy model of EH. It naturally generalises 
to include the case where the entropy of the central gas declines  
during the lifetime of the cluster, as is the case if the radiative
cooling of the ICM dominates over the shock heating that occurs during 
cluster-cluster mergers.

The structure of the paper is as follows. Section~2 develops the
entropy-driven model, firstly by introducing the concept of cluster-core
entropy and then by parameterising its evolution through the parameter
$\epsilon$. The physical processes affecting the core entropy are
discussed in Section~2.2, and revised scaling relations are discussed
in Sections 2.3 and~2.4. Finally, in Section~2.5, I describe a 
limitation to the viability of the evolutionary model.
Section~3 discusses the role of radiative cooling in the evolution
of the ICM, including a comparison with the cooling-flow driven model
of Waxman \& Miralda-Escude (1995) in Section~3.2. In Section~4, I consider
the practical problem of using realistic observational data to constrain
the allowed range of evolutionary models. Although the currently available 
data for distant clusters are relatively poor,
significant restrictions are already imposed. Furthermore, it is clear
that the definition of a unique model will be feasible in the near 
future. A summary of the conclusions of this paper is presented in Section~5.  

Throughout this paper, I assume that the growth and
collapse of cluster-scale density perturbations occurs in a 
critical density universe. In such a universe, the gravitational growth 
of clusters is described by the effective slope of the density
fluctuation power spectrum ($n$), and it is sufficient to introduce
a single parameter ($\epsilon$) to describe the evolution of the 
minimum entropy of the intra-cluster medium. This restriction
ensures that the scaling relations on which I draw can be robustly
justified, and ensures that the model parameter-space is suitably
limited. However, by suitable reformulation of the growth of cluster
properties, it is possible to 
generalise these results to the evolution of clusters in 
an open or vacuum-energy dominated universe. This work will be presented
in a subsequent paper.

\section{A Family of Entropy-Driven Evolutionary Models}

\subsection{Why introduce Entropy?}

As described by EH, the gas core radius, $r_c$, the cluster virial
radius, $r_v$, the central gas density, $\rho_c$, and the mean
background density, $\rho_b$, are related through
$$
   \left(r_c\over r_v\right)^3 \sim \left(\rho_c\over\rho_b\right)^{-1/\beta}   
\eqno(1)$$
where $\beta$ is the standard density profile parameter:
$\rho(r) = \rho_c\left(1+\left(r/r_c\right)^2\right)^{-3\beta/2}$.
This expression assumes that the gas density at the virial radius
is a constant multiple of the background density, and that the gas density
profile traces the total mass density profile in the outer parts of the 
cluster but flattens out within $r < r_c$ to form a 
core. This core is assumed to be present only in the gas distribution: 
the mass is assumed to maintain its $r^{-3\beta}$ slope to much smaller 
radii. Clearly, an equivalent of Equation~1 could also be derived for
other density profile parameterisations such as that proposed by
Navarro, Frenk \& White (1995, NFW). Since this profile
is no longer a simple power-law, it would greatly complicate the
form of the equations that follow. However, in order to model the evolution
of clusters over a limited range of redshift, it is adequate to approximate
the density profile by its local gradient (ie., 
$\beta \approx -{1\over3}{d\ln\rho\over d\ln r}$).  

In order to elucidate the physical significance of the gas core, it is 
necessary to introduce the concept of specific entropy. This is defined as
$$
	s = \cv \ln\left(T\over\rho^{\gamma-1}\right)
\eqno(2)$$
where $T$ is the gas temperature, $\cv$ is the specific heat capacity of
the gas at constant volume and $\gamma$ is the ratio
of specific heats at constant pressure and constant volume.
Since the temperature profile of the gas remains
approximately flat outside the cluster core, $s$ must fall towards the 
cluster centre. Note that even though recent ASCA results have shown that the
temperature profile is not exactly constant (eg., Markevitch et al., 1996, 
Markevitch, 1996), this statement remains true because of the very strong 
radial dependence of the gas density. 
For example, the gas density falls by a factor $\sim50$ over the range of
Markevitch et al.'s temperature measurements: if the gas were to have no 
radial entropy gradient, this would imply a factor of 14 change in 
temperature over this range. The observed factor is 3.
The core in the gas distribution corresponds to a minimum in the gas entropy:
$$
	\rho_c \approx \left( T e^{-\smin/\cv}\right)^{1/(\gamma-1)}.
\eqno(3)$$
In what follows, I will take the view that the core in the density 
distribution {\it is caused by\/} the existence of a minimum 
entropy of the ICM (an alternative interpretation is briefly
discussed in Section~3.2). Writing Equation~3 in terms of entropy and
temperature therefore separates the effects of adiabatic compression 
(increases $T$, but leaves $\smin$ unchanged), and shock heating and/or 
radiative cooling (increases [or reduces] $\smin$ with little effect on $T$). 
The last part of this statement follows from the
assertion that the global cluster temperature is proportional to the 
cluster virial temperature and the assumption that the cluster is roughly 
isothermal outside the cluster core. 

A more detailed understanding of the balance between gas entropy and 
temperature (or pressure) is clearly desirable. However, although 
spherically symmetric infall models (eg., Bertschinger, 1985) can give some 
insight into the build-up of gas in the cluster potential, they are 
inherently unrealistic since the growth of clusters is intrinsically 
hierarchical. Reliable progress can only be made through high resolution 
hydrodynamic simulations, such as those presented by NFW; but these
must be analysed with caution to ensure that the finite resolution is
fully taken into account. By focusing attention on the entropy of the gas 
only in the core of the cluster, much of the complexity of this issue can 
be side-stepped. In what follows, I treat the evolution of the core 
gas entropy as a phenomenon that is to be determined empirically.

So far, however, I have achieved little practical advance. 
EH developed this model by assuming that $\smin$ was non-evolving. 
Advances in the measurement of the X-ray evolution of clusters now justify 
a more general set of models. This paper expands EH's work by parameterising
the evolution of the central gas entropy as a power of the expansion factor: 
$$
	\smin = \smin(z=0) + \cv\epsilon\ln(1+z).
\eqno(4)$$
Each value of the parameter $\epsilon$ generates a new model for the evolution
of the X-ray properties of galaxy clusters which can be interpreted in terms 
of the physical processes responsible for the heating and cooling of the 
intra-cluster gas. With this parameterisation of the core gas entropy,
the evolution of the relative density of gas in the core can be written
$$
	{\rho_c\over \rho_b} \propto T^{1\over\gamma-1} 
		(1+z)^{-3-{\epsilon\over\gamma-1}}.
\eqno(5)$$
Equivalent expressions can be derived for other density profile slopes.
The main effect of altering the beta parameter is to change the redshift
dependence implied by a given $\epsilon$. Since, as I will argue 
in Section 4, the value of epsilon must be determined from the data itself, 
a consistent description of the cluster evolution will be obtained 
even if the true value of $\beta$ differs slightly from the value assumed.
In the discussion that follows, I fix $\beta$ at its fiducial value of
$2/3$. This provides a coherent treatment within which cluster evolution
can be discussed.

\subsection{The Physical Significance of the $\epsilon$ Parameter}

There are three important regimes for the entropy evolution parameter. 
This section briefly outlines their physical interpretation. 

The case $\epsilon < 0$
corresponds to an intra-cluster medium that is being continually heated 
in each generation of cluster collapse. This may arise purely due to the 
action of shock waves during the cluster relaxation process, but the 
injection of heat by the galaxies themselves (for example, in the form 
of supernova blast waves) may also contribute. It is extremely difficult 
to estimate the heating rate {\it ab initio}, even if shock heating is
considered alone. The simulations of NFW suggest that most of the energy
in the shock front is deposited in the outer-parts of the cluster; the
shocks reaching the central part becoming weak. This suggests that we should
expect values of $\epsilon$ close to zero. It is, however, possible that
more negative values might be found in real clusters due to the cumulative
effect of large numbers of weak shocks that are not well-resolved in the
simulations.   

One particular negative value of $\epsilon$ corresponds to the
case in which the evolution of the cluster's ICM parallels
the evolution of its dark matter potential. This is the familiar
self-similar evolution model introduced by Kaiser (1986). The value of
entropy evolution parameter required to produce self-similarity 
(ie., ${\rho_c\over \rho_b}$ constant)
depends on the spectrum of density fluctuations and the ratio of 
specific-heats. Assuming $\gamma=5/3$, as appropriate for a monatomic
gas,
$$
	\epsilon_{SS} = - \left( n+7 \over n+3 \right)
\eqno(6)$$
where $n$ is the effective spectral index of density fluctuations on
cluster scales (ie., $\delta\rho_b/\rho_b \propto r^{-(n+3)/2}$). 
For flatter power spectra, each successive scale
collapses in rapid succession and the heating of the ICM must become
stronger if self-similarity is to be maintained. For example, $n=-1$ requires
$\epsilon_{SS}=-3$; $n=-2$ requires $\epsilon_{SS}=-5$. 

As I have already described, the constant entropy model of EH corresponds
to $\epsilon=0$. This model is appropriate if the shocks that
are generated during the growth of clusters are ineffective at heating
the gas in the core of the cluster. In this model, it is possible
to interpret $\smin$ as a `primordial' entropy that was established in the 
gas before it became bound into clusters. 

In the set of models with $\epsilon > 0$, the gas in the core of the 
cluster is able to radiate a significant fraction of its internal energy over 
the Hubble time.  This radiation gives rise to 
the `cooling flows' that are well established in nearby clusters 
(eg., Fabian et al., 1991), and have recently been detected also
in distant systems (eg., Donahue \& Stocke, 1995). The likely contribution
of radiative cooling to $\epsilon$ is discussed in Section~3.

\subsection{The Connection with Cluster X-ray Luminosity}

EH show that the X-ray luminosity of a cluster is
$$
	L_X \propto T^\alpha \rho_c^2 r_c^3 
\eqno(7)$$
where the appropriate value of the exponent $\alpha$ depends on whether the
luminosity is measured with a bolometric or wide-band detector 
($\alpha \sim 1/2$) or through a low-energy band-pass ($\alpha \sim 0$) as 
would be appropriate for the ROSAT or Einstein satellites. Using Equation~1, 
the X-ray luminosity can be written in terms of the background density 
($\rho_b$, which is set by the cosmic epoch), the cluster's total mass ($M$) 
and virial temperature (both of which are properties of the dark-matter
component of the cluster), and the ratio of the cluster core density to
the background density (this is determined by the entropy of the central
gas):
$$
	L_X \propto \rho_b T^\alpha M 
		\left(\rho_c \over \rho_b\right)^{2-(1/\beta)}    
\eqno(8)$$
Combining this with the evolution of the gas core density described by
Equation~5, and relating the cluster temperature to mass through
$M \propto T^{3/2} (1+z)^{-3/2}$ gives 
$$\eqalign{
	L_X \propto \, & (1+z)^{3/2 - (3+{\epsilon\over\gamma-1})(2-{1\over\beta})}\cr 
	&\qquad         T^{\alpha + 3/2 + ({1\over\gamma-1})(2-{1\over\beta})}\cr
	}
\eqno(9)$$
The appearance of this relation is substantially improved by setting 
$\gamma=5/3$ and restricting attention to cluster profiles with $\beta=2/3$:
$$
	L_X \propto (1+z)^{-3\epsilon/4} T^{(9/4)+\alpha}
\eqno(10)$$

\subsection{Strong versus Weak Self-Similarity}

So far, I have not needed to be careful about the exact meanings of the 
terms used in (eg.) Equation~10. In principle, $T$, $L_X$, $M$ etc., apply to
individual clusters, but it is by no means clear that the constant
of proportionality that links them should be the same for all clusters.
For instance, studies of the growth of hierarchical clusters (eg., Lacey \& 
Cole, 1993) show that a great variety of trajectories may lead to 
an individual cluster of given mass at the present epoch. To give an explicit 
example, we should more correctly interpret the redshift ($z$),
appearing in Equation~10, as the epoch at which the cluster last had 
a major merger event rather than the epoch at which the cluster is observed
(cf., Kitayama \& Suto, 1996).

The Weak Self-Similarity principle (eg., Kaiser, 1986) asserts that, although
there is considerable scatter between the formation histories of individual
clusters, relations of the form of Equation~10 can be used to relate the
characteristic properties of cluster populations at one cosmic epoch to 
those at another. The reasoning behind this is that --- so long as the 
power spectrum of density fluctuations contains no additional scale ---
the only physical scale that distinguishes the properties of clusters
at one epoch from those at another is the scale at which average density fluctuations
make the transition between linear and non-linear growth. In order
that Equation~10 then fits into this scheme, we must be careful also
to interpret Equation~4 as applying to the characteristic central entropy
of the cluster population (where we may either consider the cluster population
as a whole, or limit ourselves to a well-defined relative sub-population
such as the most massive 10\% of the population).
 
The form of Equation~10 has been chosen to separate the evolution
of the clusters' X-ray luminosity that comes from the increasing 
characteristic density of the universe, and the component that comes
from the changing characteristic temperatures of clusters.
Adopting a particular form for the slope of the density fluctuation
power spectrum, I arrive at the scaling relation
$$
	{\Delta\log L_X \over \Delta\log(1+z)}= 
			\left(\alpha\left(n-1\over n+3\right)
			+ {9-3\epsilon\over4} - {9\over n+3}\right).
\eqno{(11)}$$
This illustrates the balance of terms tending to increase and
decrease the luminosity as a function of redshift. Flatter power
spectra (more negative $n$) and more positive entropy evolution 
both tend to reduce the X-ray luminosities of distant clusters. 
This is balanced by an intrinsic increase in the efficiency of X-ray
emission due to the high average density of the universe.  Figure~1
illustrates this interplay between the power spectrum and entropy
evolution by applying the scaling of Equation~11 to the low-redshift
luminosity function of Edge et al., 1990. More precise definition of the 
local luminosity function will shortly be available from the ROSAT All-Sky 
Survey (eg., Ebeling et al., 1996); however, because of the degeneracy
seen in this figure, it is difficult to interpret the evolution of
the X-ray luminosity function in terms of a single combination of $n$ 
and~$\epsilon$. The two effects become easier to discriminate, however, 
as a greater range of X-ray luminosities is probed 
(cf., Castander et al., 1995).

\beginfigure{1}
\psfig{figure=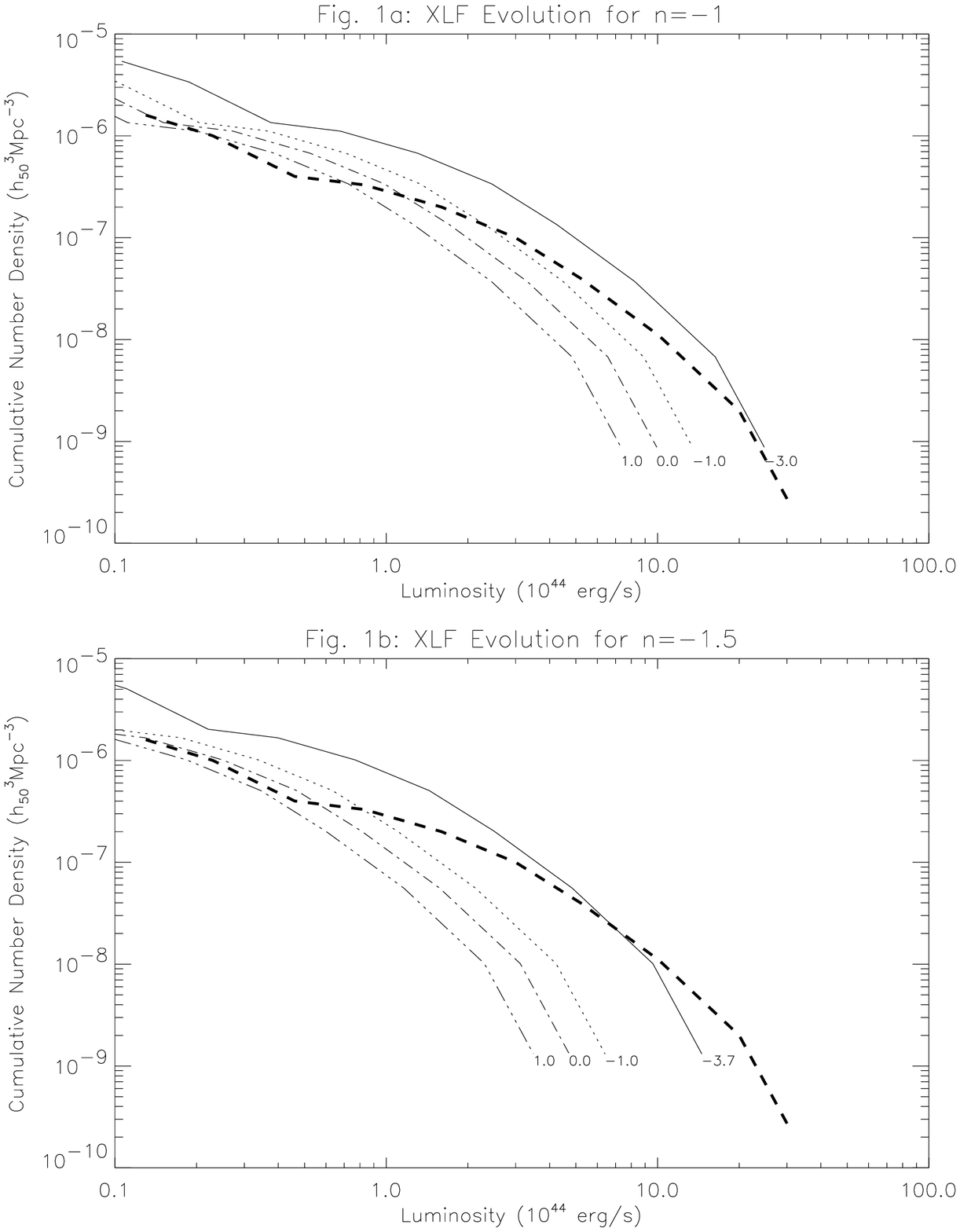,width=8.4cm}
\caption{{\bf Figure 1a, b.} The evolution of the X-ray luminosity function.
The dashed line shows the present-day cumulative X-ray luminosity function of
Edge et al., 1990.  The evolution of the gravitational potential
is described by self-similar evolution with power spectra of $-1$
(panel a) and $-1.5$ (panel b). The solid lines show the $z=0.5$ luminosity
function predicted for self-similar evolution of the ICM
($\epsilon=\epsilon_{SS}$, Equation~6); milder heating of the ICM,
$\epsilon=-1$, is shown by the dotted line; dot-dashed lines show the 
luminosity evolution for constant core entropy, $\epsilon=0$; and 
dot-dot-dashed lines illustrate a model in which gas cooling dominates
the entropy evolution, $\epsilon=+1$. All three calculations
use $\alpha=0.5$ and $\beta=2/3$.}
\endfigure

Writing Equation~10 to separate the effects of redshift and temperature
tempts another approach. The form of the relation appears to suggest
that the slope of the present-day $L_X$--$T$ correlation should be 
$\alpha +9/4$. As noted by Evrard \& Henry, this is 
encouragingly close to the observed $L_X$--$T$ slope 
($L_X^{\rm bol} \propto T^3$, eg., Edge \& Stewart, 1991, David et al., 
1994). However, this prediction goes far beyond the weak self-similarity 
principle that I have described above:
it requires that I apply the scalings not to the characteristic properties
of the clusters at a particular epoch, but to interrelate the properties
of clusters at the same epoch. This is the so-called Strong Self-Similarity
Principle. For the reasons I have described above, in particular the
different average formation histories of high and low-mass clusters,
it has little physical basis. To illustrate this point, the slope
of the $L_X$--$T$ appears to be in `agreement' with the observations even 
if $\epsilon = \epsilon_{SS}$ (Equation~6). This contrasts with the standard
self-similar model (eg., Kaiser, 1986) that `predicts' 
$L_X \propto T^{(3/2)+\alpha}$ at a fixed epoch. The discrepancy is not real,
and has arisen because the Strong Self-Similarity Principle requires that 
$\smin$ in Equation~4 applies to all clusters regardless of their mass,
rather than to the characteristic entropy of the cluster population.
This is an additional assumption for which there is no prior physical 
justification.

Although Equation~9 cannot therefore be used to estimate the slope of the 
present-day $L_X$--$T$ correlation robustly, it does predict how the 
normalisation of the correlation will evolve. Because the slope of the observed
relation is close to the temperature dependence of Equation~10,
the evolution of this relation is dominated by $\epsilon$ (Figure~2).
This breaks the degeneracy between $\epsilon$ and $n$ inherent in luminosity 
function measurements. Thus, by combining the measurement of both the
$L_X$--$T$ correlation and the luminosity function at high redshift, the 
separate roles of the power-spectrum and the core gas entropy can be 
distinguished without having to measure the evolution of the temperature 
function directly. The practical feasibility of this approach is discussed 
in Section~4.

\beginfigure{2}
\psfig{figure=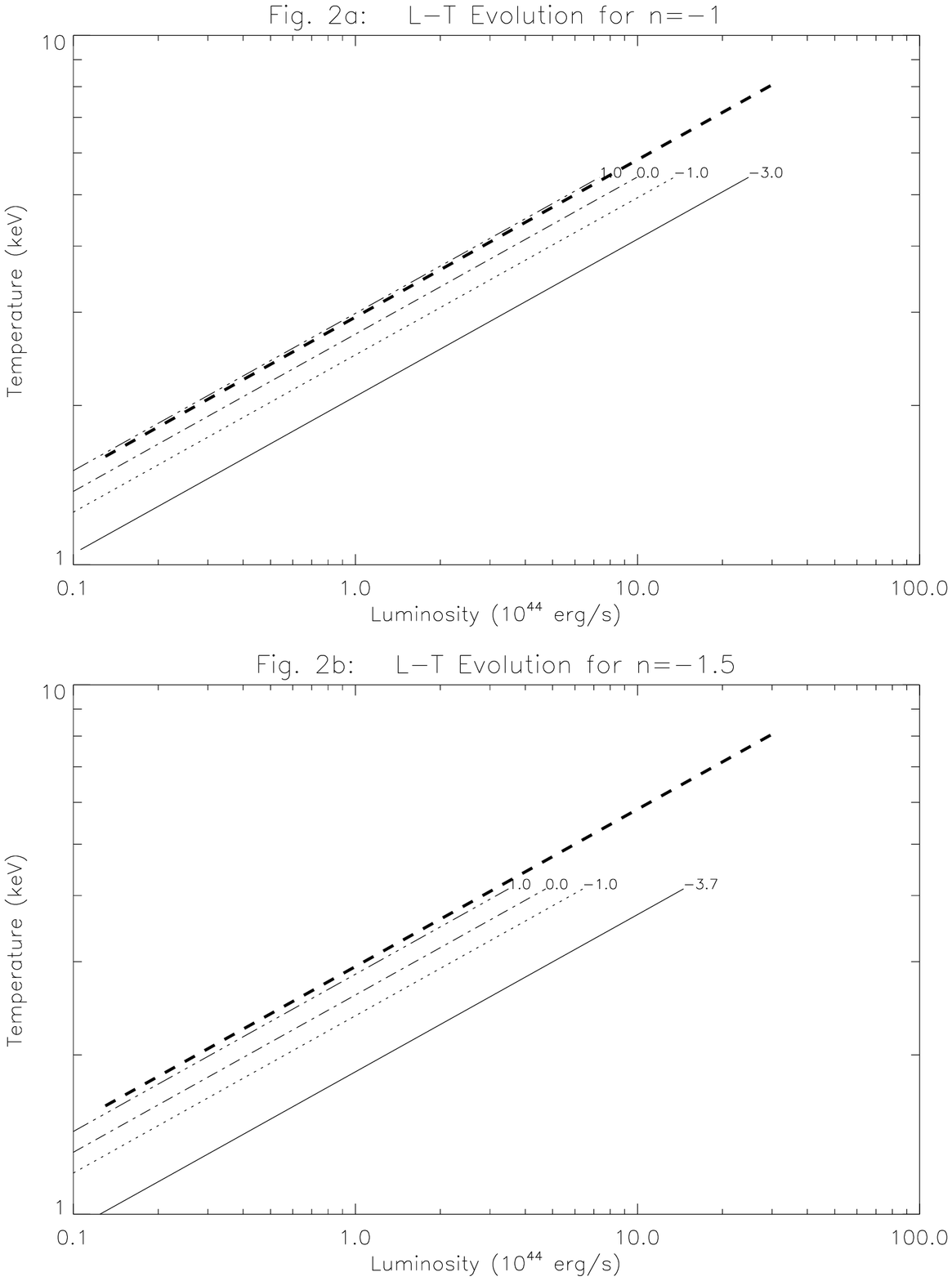,width=8.4cm}
\caption{{\bf Figure 2a, b.} The predicted evolution of the X-ray 
luminosity -- temperature correlation. Panel a shows $n=-1$; panel b, 
$n=-1.5$. The thick dashed line shows the position of the present-day 
correlation from David et al., 1993 (as appropriate for bolometric 
luminosities). The other line types illustrate the effect of different 
choices of the $\epsilon$ parameter on the correlation at $z=0.5$. 
The evolution of the correlation is only weakly affected by the choice of 
density fluctuation spectral index, $n$. This most strongly affects the 
range of cluster temperatures that will be observed. As in Figure~1, I
have used $\alpha=0.5$ and $\beta=2/3$.}
\endfigure

\subsection{Limit on the Validity of the Model}

The model that I have described implicitly assumes that the core radius
is set by the minimum entropy of the cluster gas remains well inside
the cluster's virial radius. This corresponds to the standard picture
of a cluster in which the gas is confined by a virialised dark matter
halo.

The scaling of the ratio of the core and virial radii is given by
$$
	{\Delta\log\left(r_c\over r_v\right) \over \Delta\log(1+z)}
		=  {1\over3\beta(\gamma-1)} \left[ 3(\gamma-1)
		+ \epsilon - \left(n-1\over n+3\right) \right]
\eqno{(12)}$$
Adopting a typical present-day ratio of 1/10 for $r_c / r_v$, and setting 
$\beta$ and $\gamma$
to their standard values, the epoch at which the gas is no longer trapped 
within the virial radius of a typical halo is
$$
	\log(1+\zmax) = \left[{3\over4}
			\left(\epsilon+{n+7\over n+3}\right)  \right]^{-1}
\eqno{(13)}$$   
This limit becomes more problematic for larger values of $\epsilon$
and flatter (more negative) power spectra. For example, 
$(n,\epsilon)=(-1,+1)$ has $\zmax=1.2$ while $(n,\epsilon)=(-2,+1)$ has 
$\zmax=0.67$. Even in these cases, however, the model still provides a 
framework for understanding the evolution of galaxy
clusters over the range of redshifts that are currently observable.
Furthermore, when $r_c \sim r_v$, the scaling relations 
may continue to apply if the cluster's gravitational potential is 
sufficiently smooth and well-defined in the infall region.


\section{The Role of Radiative Cooling in Cluster Evolution}
  
\subsection{Estimating the Evolution of Gas Entropy due to Cooling}

During cluster mergers, dissipation of the bulk motion of the gas tends
to raise the specific entropy of the ICM. Although it is difficult to estimate 
the effectiveness of this process for core gas, it is clear that we should 
expect $\epsilon < 0$. The radiative cooling of the intra-cluster gas, 
which occurs principally at X-ray wavelengths, acts in the opposite sense.
In this section, I use the observed temperatures and densities of gas in the 
cores of present-day clusters to estimate an effective value for 
$\epsilon_{\rm cool}$, the radiative cooling contribution to the $\epsilon$ 
parameter. In the absence of alternative cooling mechanisms, this estimate is 
an upper limit to the value of $\epsilon$ that will actually occur.

The change in specific entropy with time due to radiative cooling is given by
$$
	{ds\over dt} = - {K_L n_e \over T \mu_H m_H}
\eqno(14)$$
where $K_L$ is the cooling coefficient (ie., the radiated power per unit 
volume is $K_L n_e n_H$), $n_e$ is the electron number density, $\mu_H$ is
the relative atomic mass of the plasma per hydrogen atom, $m_H$ is the 
atomic mass unit, and $T$ is the gas temperature in Kelvin. For a cosmic 
abundance plasma with temperature $\sim 5\times 10^7K$, eqn.\ 14 can be 
written
$$\eqalign{
	{ds\over dt} = & - 1.8\times10^{-10} 
			\left[n_e\over10^{-3}\cm^{-3}\right]
			\left[K_L\over2\times20^{-23}\ergs\cm^3\right]\cr
		& \qquad\times  \left[T\over5\times10^7\,\hbox{K}\right]^{-1}
			\ergs\hbox{K}^{-1}\hbox{g}^{-1}\cr
	}
\eqno{(15)}$$
For such a plasma, the specific heat capacity is 
$$
	\cv = {{3\over2}k_B \over \mu m_H} = 2.07\times10^8
		\,\hbox{erg}\,\hbox{K}^{-1}\hbox{g}^{-1}
\eqno{(16)}$$
where $\mu$ is the mean molecular weight per free particle.
Using the definition of $\epsilon$ given in Equation~4, and transforming 
Equation~14 to a derivative with respect to $1+z$, the effective value of the 
dimensionless entropy parameter can be expressed as:
$$\eqalign{
	\epsilon_{\rm cool} = & 0.53\,(1+z)^{-3/2} h_{50}^{-1} 
			\left[n_e\over10^{-3}\cm^{-3}\right]\cr
	& \qquad \times \left[K_L\over2\times20^{-23}\ergs\cm^3\right]
			\left[T\over5\times10^7\,\hbox{K}\right]^{-1}\cr
	}
\eqno{(17)}$$ 
where the Hubble constant is parameterised by $H_0 = 50\,h_{50}\kms\Mpc^{-1}$. 

The redshift dependence of $\epsilon_{\rm cool}$ can be estimated
by combining Equations 5 and~7. Adopting $\gamma = 5/3$, 
$\epsilon_{\rm cool}$ varies as $(1+z)^{-3(1+\epsilon)/2} T^{1/2}$,
so that even if the low redshift evolution is dominated by cooling,
it becomes less and less important as we look to higher redshifts.

By extracting from the literature values for $T$ and $n_e$ for gas in the 
centres of present-day clusters of galaxies, it is possible to estimate the 
appropriate values of $\epsilon$ for specific systems. These will serve as a 
guide to the value of $\epsilon$ that applies to the cluster population of as 
a whole. For the Coma cluster (Briel et al., 1991), I find 
$\epsilon_{\rm cool}=0.80$; for the extreme cluster A2163, 
$\epsilon_{\rm cool}=1.5$ (Markevitch et al., 1996) [both calculations
assume $h_{50}=1$].
As I have stressed, the actual value of $\epsilon$ that describes the X-ray
evolution of galaxy clusters is likely to be less than 
$\epsilon_{\rm cool}$ due to the heating of core gas during cluster formation. 
However, the analysis of this section shows that we can conservatively 
expect that $\epsilon < 2$ at the present-day, with the limit becoming even 
more stringent at higher redshifts. This sets an important upper
limit of the strength of the decline in the X-ray luminosities of clusters.

\subsection{Comparison with the Cooling Flow-Driven Model of 
	Waxman \& Miralda-Escude}

Waxman \& Miralda-Escude (1995, WM-E) present a model for the evolution
of spherical clusters in which the core in the gas density profile is set by 
the surface (the `cooling radius') 
at which the cooling time of the gas equals the age of
the universe at the epoch under consideration. The evolution of
the gas core density with redshift is thus set by the competition between the 
increasing cooling efficiency (due to higher characteristic densities)
and the falling age of the universe. For a $\beta=2/3$ cluster profile,
they find
$$
	{\rho_c\over\rho_b} \propto T^{1-\alpha'} (1+z)^{-3/2}
\eqno{(18)}$$
where $\alpha'\approx1/2$ parameterises the temperature dependence of the 
{\it bolometric} X-ray luminosity (as opposed to the luminosity measured
by a band-pass detector). Comparison with Equation~5 shows that
the cooling flow model can be formally incorporated into the entropy-driven 
model by setting $\gamma = 1 + 1/(1-\alpha') \approx 3$ and 
$\epsilon = -3 / 2 (1-\alpha') \approx -3$. The value of $\gamma$
required does not, however, have a physical interpretation: the temperature
dependence of the density ratio is considerably weaker than in the
standard $\gamma=5/3$ entropy-driven model. 

The evolution of X-ray luminosity in WM-E's model is given by
$$
	L_X \propto T^{2+\alpha-(\alpha'/2)} (1+z)^{3/4}.
\eqno(19)$$
Comparing this with Equation~10 shows that, compared to an $\epsilon>-1$
entropy driven model, the cooling flow model will produce more rapid 
increase (with redshift) in X-ray luminosity at a given temperature. 
Although in both cases the ratio of the gas core radius to the virial radius 
increases with redshift, at a given temperature, the strength
of this effect is larger for the $\epsilon>-1$ model. Thus in the latter class
of model, an ever smaller fraction of the ICM becomes susceptible to
the cooling flow instability. Fortunately, observation of
the evolution of the $L_X$--$T$ correlation will provide a simple test by which
the two classes of model can be distinguished. Only the entropy-driven
model is able to account for a relation that is non-evolving or in 
which the X-ray luminosity at a given temperature falls with redshift.

Although WM-E's cooling flow driven model cannot naturally be expressed
in terms of a particular value of $\epsilon$, the general nature of the
models presented here guarantees that the observational signature of
such models can be reproduced by an apparent combination of $n$ and
$\epsilon$ parameters. I describe the parameters as `apparent' since
although they reproduce the observations, they will not correspond
to true spectral index, or describe the true evolution of cluster
central entropy. Comparing equations 19 and 10 shows that relation between 
the true and apparent spectral index is given by 
$ {n_{ap}-1 \over n_{ap}+3} = {7\over9}\left(n-1\over n+3\right)$ (for example,
a true spectral index of -1.0 would give an apparent index of -0.75)
and that all WM-E models will have $\epsilon_{ap} = -1$. In the following
section, I will examine how realistic data may be used to constrain
the true cluster evolution model in ($n$,$\epsilon$) parameter space. 
It must, however, be clearly stated that a measurement of $\epsilon\approx -1$ 
may be equally well interpreted within the entropy-driven model described
in this paper, or within the cooling-flow driven model described by WM-E.

\section{Observational Constraints on the Parameters \hbox{$n$} and $\epsilon$}

Measurement of the evolution of the galaxy cluster temperature function
would provide a clean test of the rate of cluster mass growth and thus 
allow accurate determination of the density fluctuation spectral 
index effective on cluster scales (eg., Eke et al., 1996). An idealised 
measurement would be made from a cluster survey in which selection was based 
on temperature and was independent of cluster X-ray flux or luminosity. 
In practice, such a survey would have to be constructed from a flux limited 
cluster sample. One strategy would be to discard enough clusters to create 
a volume limited subsample, to measure the temperatures of these clusters and 
then to apply a second threshold so that the sample became complete in 
temperature. A clearly preferable procedure would use two data-sets to 
define separately the evolution of the X-ray luminosity function (XLF) and 
the evolution of the X-ray luminosity -- temperature (L--T) correlation. 
Within the entropy-driven model,
these two pieces of data create near-orthogonal constraints on the 
parameters $n$ and $\epsilon$ thus allowing us to piece together the 
complete picture of cluster evolution. While the two data-sets must refer 
to clusters of similar luminosities, there is no requirement that the
L--T correlation be determined from a statistically complete sample.

Figure 3 illustrates the constraint that can be placed on the values of $n$
and $\epsilon$ for a range of hypothetical evolutionary measurements. 
Throughout this section, I have assumed that the density profile can be 
adequately approximated by setting $\beta=2/3$.
Solid lines illustrate the constraint from the evolution of the XLF. The 
figure assumes that the cumulative XLF can be approximated by a power-law 
of slope $-1.2$ over the relevant range of X-ray luminosities and sets 
$\alpha=0$, as appropriate for a low energy band-pass detector such as 
ROSAT. The lines are labeled by the logarithm of change in the XLF 
amplitude (measured at a fixed X-ray luminosity) between the present-day 
and $z=0.5$.
For example, the line labeled 0 shows the constraint implied by a 
non-evolving luminosity function; that labeled by $-0.3$ is the constraint 
implied by a factor of 2 fall in XLF amplitude. Dashed lines in the figure
show the equivalent constraint implied by a measurement of the evolution
of the L--T correlation. The labels give the logarithm of the change in the 
normalisation (ie., temperature at a fixed X-ray luminosity) between the 
present-day and $z=0.5$. An L--T inverse slope of $0.30$ is assumed, and the 
luminosities are taken to be bolometric (ie., $\alpha=0.5$). 

\beginfigure*{3}
\psfig{figure=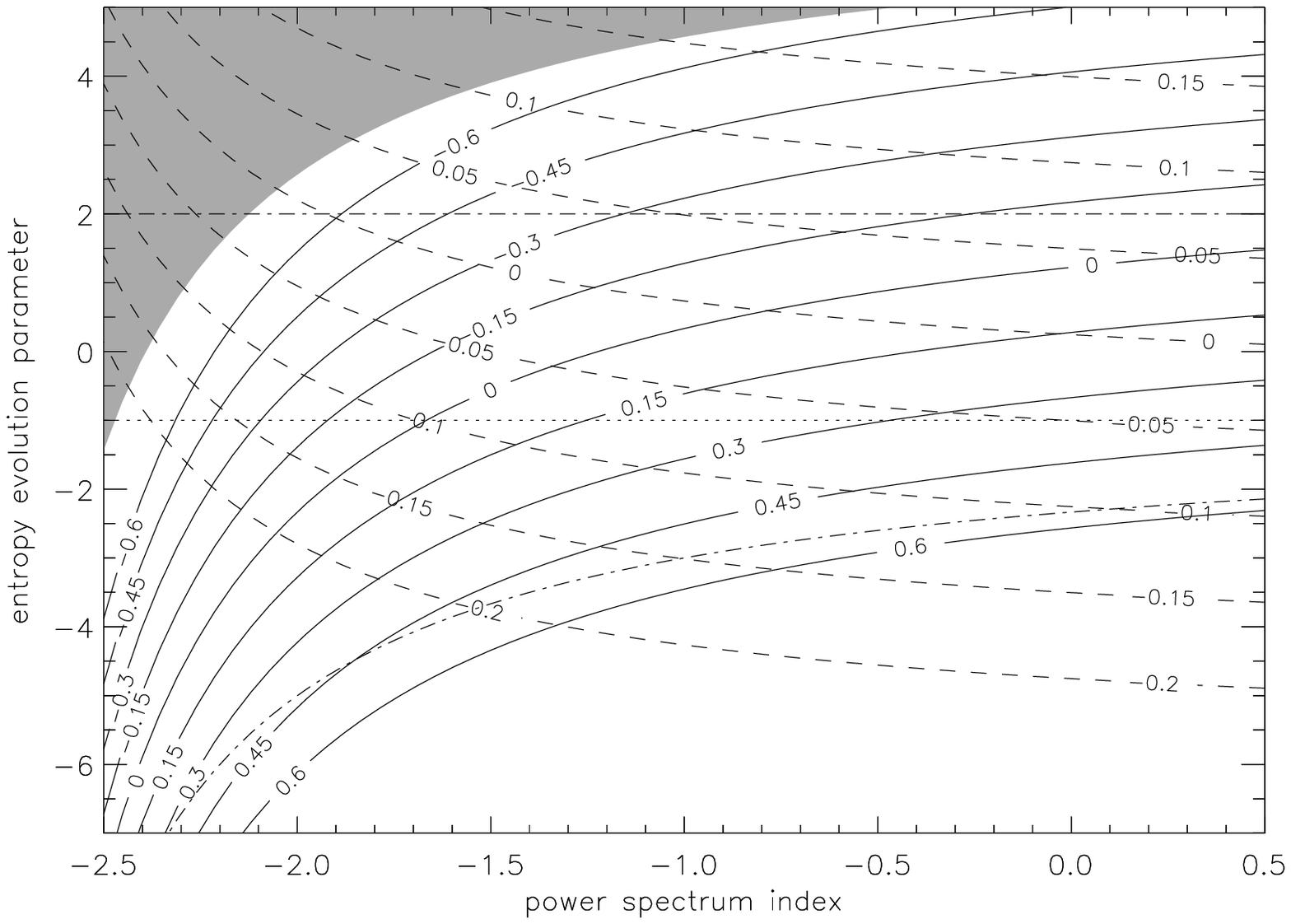,width=16.8cm}
\caption{{\bf Figure 3.} Constraints on the parameters $n$ and
$\epsilon$ inferred from the evolution of the X-ray luminosity function
(solid lines) and the evolution of the luminosity -- temperature
relation (dashed lines) as it might be measured at redshift 0.5. 
The lines are labeled by the logarithm of the change in normalisation: 
further details are given in the text of Section~4. 
Dot-dashed lines indicate values of the 
$\epsilon$ parameter corresponding to self-similar evolution 
(bottom line) and evolution dominated by radiative cooling (upper line);
while the dotted line shows parameter space occupied by the
cooling-flow driven models of WM-E. The shaded area indicates the
region of parameter space in which the underlying assumptions of our
model are no longer valid at redshift 0.5.}
\endfigure

Any single evolutionary measurement (of either the XLF or the L--T correlation)
can be reproduced by a wide range of ($\epsilon$,$n$) parameter pairs. However,
measurement of the evolution of both observable relations removes this
degeneracy. For flat power spectra (ie., more negative $n$), the lines of given
XLF and L--T evolution are nearly orthogonal. Thus in this region even
low accuracy measurements can place stringent constraints on $n$ and
$\epsilon$. As $n$ becomes more positive, the rate of gravitational growth
slows down and the evolution of both the XLF and L-T relations is driven 
by entropy evolution alone. In this region of the diagram, the two 
measurements become more degenerate and greater accuracy is required to 
determine $n$.       

The figure also gives two dot-dashed lines illustrating the limiting physical
 models. The lower line shows the curve defined by self-similar models, 
$\epsilon = \epsilon_{SS}$. Unless gas is heated more than the dark matter 
during cluster collapse, the evolutionary model must lie above above this 
line. The upper line shows the maximum evolution due to radiative cooling 
($\epsilon = 2$) suggested by the analysis of Section~3. Unless some other 
cooling mechanism can be invoked, or the value of $\beta$ is significantly
different from $2/3$, an acceptable model must lie below this line.
In addition, the shaded upper left-hand corner illustrates the area of
parameter space for which $z_{\rm max}$ (as defined by Equation~13)
is less than $0.5$. The underlying assumptions of the model are no longer
valid in this region. The shaded region would become larger if we were
to trace the evolution to  still higher redshifts. Finally, I have added
a dotted line to the figure to show the area of (apparent) parameter 
space occupied by the models of WM-E.  
 
\beginfigure*{4}
\psfig{figure=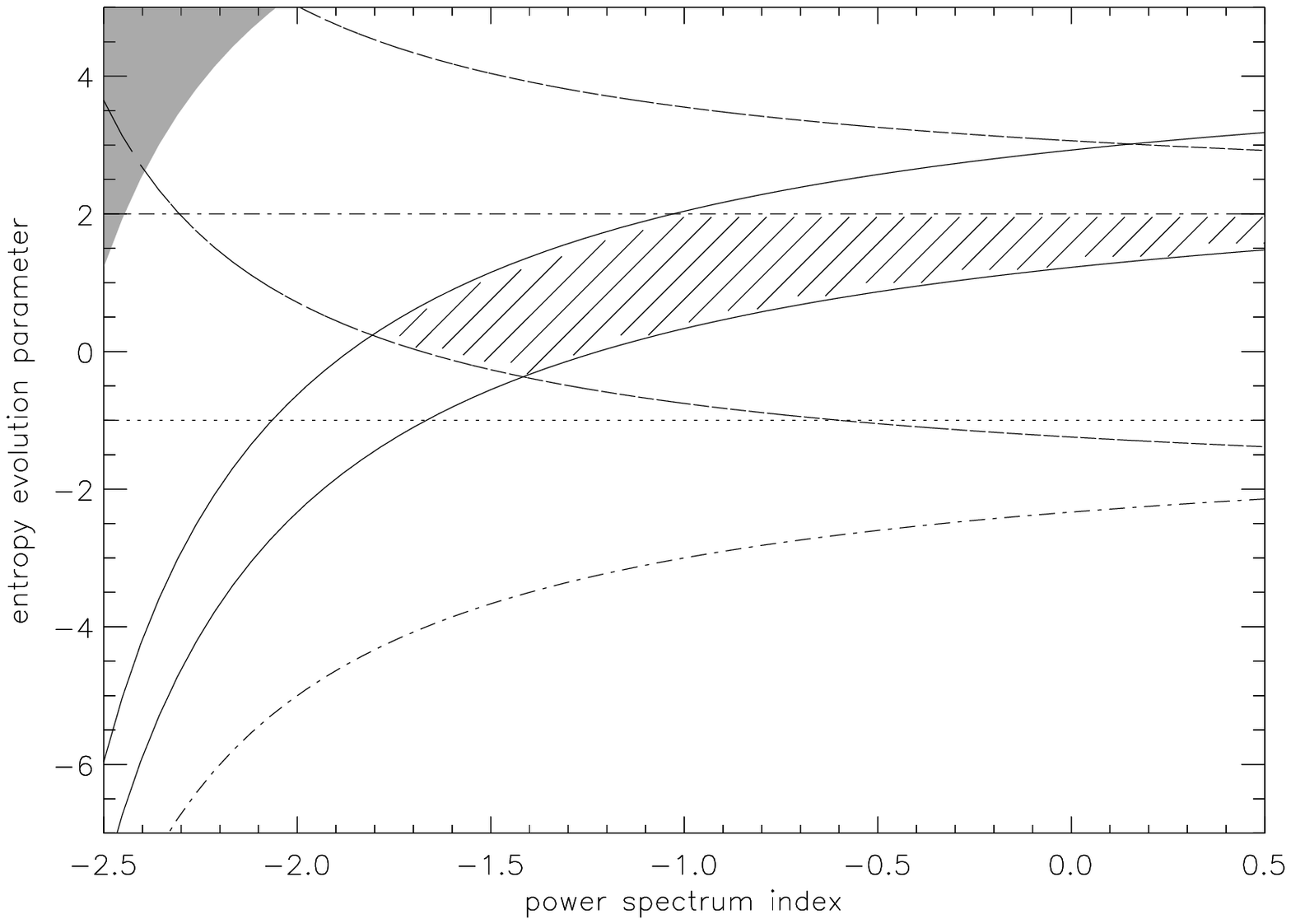,width=16.8cm}
\caption{{\bf Figure 4.} Preliminary constraints on the cluster
evolutionary model from current data for distant clusters. 
Solid lines show the present limits of the X-ray luminosity function 
evolution at $\bar z=0.35$ from Castander et al., 1995 (upper line), and 
Nichol et al., 1996 (lower line). Dashed lines illustrate the 
formal 90\% confidence limits derived from the $\bar z=0.21$ 
luminosity--temperature correlation of preliminary ASCA data
(Tsuru et al., 1996). Acceptable models (indicated by the hashed region)
must lie between these lines and below the limit $\epsilon<2$.
The evolutionary calculations assume $\alpha=0.0$ for the luminosity 
function data and $\alpha=0.5$ for the luminosity--temperature correlation
data. 
Dotted and dot-dashed lines are carried over from Figure~3; the shaded
corner of the plot indicates the area of parameter space in which the
evolutionary model is invalid at $z=0.35$.}
\endfigure

Although current data leave the evolution of the XLF and the L-T correlation 
poorly defined, it is already interesting to investigate the limits that are 
implied. The situation is shown in Figure~4, where solid lines show
the constraint from luminosity function measurements. Dashed lines show the
constraints from the temperature function. The comparison is made at a 
nominal luminosity of 
$10^{44}\ergs$. To define the lower limit to evolution of the XLF, I take
the recent reanalysis of the EMSS survey by Nichol et al., 1996. Using
deeper ROSAT images of a number of the EMSS clusters, they conclude that a
non-evolving XLF cannot yet be excluded. This forms the lower solid line. 
The upper line shows a 40\% fall in XLF amplitude as suggested by Castander
et al., 1995, from the analysis of serendipitous ROSAT fields.
Both measurements are taken to have a median redshift of 0.35 and have been
normalised using the low redshift luminosity function of Henry et al.\ 
(1992).  Data for the evolution of the L--T relation are taken from 
Tsuru et al., 1996 (and normalised using $z<0.1$ data from David et al.,
1993). It should be noted that Tsuru et al.'s results are based on 
{\it preliminary} reduction of data from the ASCA satellite and should 
therefore be treated with caution. Furthermore,
the data have a median redshift of only $0.21$, so the evolutionary
constraint that can be derived from their data has relatively little
leverage. The  two curves show a change in logarithmic normalisation
of between $-0.028$ and $+0.053$ (ie., an uncertainty of $\pm 10$\% in the 
distant cluster L--T correlation amplitude), corresponding to the 90\%
formal confidence limits of Tsuru et al.'s data. 

A cursory inspection of the figure is disappointing: the density fluctuation
index $n$ can have any value grater than $-1.7$. Yet even these data make
a number of points clear. Firstly the data are inconsistent with any very flat
spectral index. While values of $n$ around $-2$ are able to reproduce the
observed XLF evolution, they produce too much evolution in the L--T relation
to be acceptable. It is clear that none of the self-similar models 
can account for the observed evolution; but the data suggest
that the combined XLF and L--T data permit only models in which gas
cooling dominates over heating. Furthermore, although these
preliminary data provide only a weak constraint on the spectral index, 
the figure shows that even a modest improvement in the
accuracy of the distant cluster L--T relation, or the median redshift
at which it can be determined, will considerably refine the determination
of $n$.

\section{Conclusions}

The basic aim of this paper has been to produce a general framework
describing the evolution of the X-ray properties of galaxy clusters.
In contrast to previous work, the model I have presented explicitly
differentiates between the contribution from the evolution of the
gravitational potential, and the component due to the evolution of
the entropy of the cluster core gas. This division of the evolution
into two components ensures that each realisation of the model then 
has a physical interpretation: the evolution of the gravitational component
is derived from the density fluctuation power spectrum and the
background cosmological model, while  
the evolution of the core entropy is determined by the balance
between the radiative cooling of the intra-cluster medium and its shock
heating during cluster-cluster mergers. 

In this paper, I have limited attention to the evolution of clusters in a
critical density universe dominated by collisionless dark matter and have 
assumed that the density fluctuation power spectrum can be represented by 
an effective spectral index $n$. 
These two assumptions allow the evolution of the cluster gravitational 
potential to be described using the self-similar scaling relations of 
Kaiser (1986). In contrast, the evolution of the gas distribution is 
determined by the assumption that the gas has a roughly isothermal 
distribution with temperature proportional to the cluster virial temperature, 
and with its core radius set by a minimum entropy constraint. The evolution 
of cluster X-ray properties is then determined, on the one hand, by the 
evolution of the cluster virial temperature, and on the other, by the 
evolution of the central entropy. The entropy evolution is specified by 
the parameter $\epsilon$ (Eqn.\ 4), which I have regarded as a quantity to be 
fixed by observation. 

The parameters $n$ and $\epsilon$ generate a two-dimensional continuum 
of scaling relations that allow the characteristic properties of clusters
at one epoch to be transformed to another. Because the effects of changes
in these parameters are close to orthogonal, they allow a very
general set of evolutionary scenarios to be generated. Furthermore, 
each point in the $n$--$\epsilon$ plane has a unique physical interpretation. 
Some regions of the parameter space can be singled out as having particular 
interest. 
One such region is the line of models in which the heating of the ICM
during the gravitational growth of the cluster maintains a constant ratio
between the cluster core and virial radii. This subset of parameters 
recovers the original self-similar scaling relations proposed by 
Kaiser (1986). Another notable dividing line is the upper limit on the entropy
evolution parameter that can be set  from the observed X-ray 
surface brightnesses of present-day cluster cores. If gas cooling occurs 
radiatively, $\epsilon$ cannot be larger than 2.  

The model described here can, however, only be used 
to derive scaling relations (ie., to use the properties of clusters
observed at the present-day to predict the properties of the cluster
population at an earlier time). For this reason, I have been careful to
avoid discussion of the way in which the central ICM entropy may vary
between clusters of differing masses at a single epoch. This question
cannot be addressed without introducing further parameters into
the model.   
 
In the final section of the paper, I have illustrated how the observed 
evolution of cluster properties may be used to discriminate between 
different models.
Although knowledge of the evolution of the X-ray luminosity function
alone does not provide enough information to determine $n$ or $\epsilon$
uniquely, accurate differentiation is possible if the 
luminosity function is combined with data on the evolution of the luminosity
-- temperature correlation. For flatter power spectra, (ie., more negative
$n$) the XLF and L--T constraints impose near orthogonal limits in the
$n$--$\epsilon$ parameter space. Thus even quite poor measurements  
result in significant constraints on the physical
model (ie., the rate of gravitational growth and rate of gas heating/cooling)
underlying cluster evolution. As a result, definition of this model is
easily within the scope of current and near-term X-ray missions.

\section*{Acknowledgments}

I would like to acknowledge the comments and encouragement of Shaun Cole,
Carlos Frenk and Gus Evrard and to thank the anonymous referee for
the improvements suggested. This project was carried out using computing 
facilities supplied by the Starlink Project, and was supported by a PPARC
rolling grant for ``Extragalactic Astronomy and Cosmology at Durham''.

\bye

\section*{Figure Captions}

\noindent {\bf Figure 1a, b.} The evolution of the X-ray luminosity function.
The dashed line shows the present-day cumulative X-ray luminosity function of
Edge et al., 1990.  The evolution of the gravitational potential
is described by self-similar evolution with power spectra of $-1$
(panel a) and $-1.5$ (panel b). The solid lines show the $z=0.5$ luminosity
function predicted for self-similar evolution of the ICM
($\epsilon=\epsilon_{SS}$, Equation~6); milder heating of the ICM,
$\epsilon=-1$, is shown by the dotted line; dot-dashed lines show the 
luminosity evolution for constant core entropy, $\epsilon=0$; and 
dot-dot-dashed lines illustrate a model in which gas cooling dominates
the entropy evolution, $\epsilon=+1$. All three calculations
use $\alpha=0.5$ and $\beta=2/3$.

\noindent {\bf Figure 2a, b.} The predicted evolution of the X-ray 
luminosity -- temperature correlation. Panel a shows $n=-1$; panel b, 
$n=-1.5$. The thick dashed line shows the position of the present-day 
correlation from David et al., 1993 (as appropriate for bolometric 
luminosities). The other line types illustrate the effect of different 
choices of the $\epsilon$ parameter on the correlation at $z=0.5$. 
The evolution of the correlation is only weakly affected by the choice of 
density fluctuation spectral index, $n$. This most strongly affects the 
range of cluster temperatures that will be observed. As in Figure~1, I
have used $\alpha=0.5$ and $\beta=2/3$.

\noindent {\bf Figure 3.} Constraints on the parameters $n$ and
$\epsilon$ inferred from the evolution of the X-ray luminosity function
(solid lines) and the evolution of the luminosity -- temperature
relation (dashed lines) as it might be measured at redshift 0.5. 
The lines are labeled by the logarithm of the change in normalisation: 
further details are given in the text of Section~4. 
Dot-dashed lines indicate values of the 
$\epsilon$ parameter corresponding to self-similar evolution 
(bottom line) and evolution dominated by radiative cooling (upper line);
while the dotted line shows parameter space occupied by the
cooling-flow driven models of WM-E. The shaded area indicates the
region of parameter space in which the underlying assumptions of our
model are no longer valid at redshift 0.5.

\noindent {\bf Figure 4.} Preliminary constraints on the cluster
evolutionary model from current data for distant clusters. 
Solid lines show the present limits of the X-ray luminosity function 
evolution at $\bar z=0.35$ from Castander et al., 1995 (upper line), and 
Nichol et al., 1996 (lower line). Dashed lines illustrate the 
formal 90\% confidence limits derived from the $\bar z=0.21$ 
luminosity--temperature correlation of preliminary ASCA data
(Tsuru et al., 1996). Acceptable models (indicated by the hashed region)
must lie between these lines and below the limit $\epsilon<2$.
The evolutionary calculations assume $\alpha=0.0$ for the luminosity 
function data and $\alpha=0.5$ for the luminosity--temperature correlation
data. 
Dotted and dot-dashed lines are carried over from Figure~3; the shaded
corner of the plot indicates the area of parameter space in which the
evolutionary model is invalid at $z=0.35$.

\bye